\documentclass[aps,prb,twocolumn,amsmath]{revtex4}

\usepackage{breqn}
\usepackage{bm}
\usepackage{graphicx}
\usepackage{cleveref}
\usepackage{natbib}

\newcommand{\nn}{\nonumber}
\newcommand{\ts}{\textstyle}
\newcommand{\ds}{\displaystyle}
\newcommand{\tr}{\,\mathrm{tr}}
\renewcommand{\ln}{\mathrm{ln}}

\newcommand{\A}{ \mathcal{A} }

\makeatletter
\let\cat@comma@active\@empty
\makeatother

\begin{document}

\title{
Nonlinear microwave response of clean superconducting films}
\author{A.A. Radkevich$^1$ and A.G. Semenov$^{1.2}$
}
\affiliation{$^1$I.E.Tamm Department of Theoretical Physics, P.N.Lebedev Physical Institute, 119991 Moscow, Russia\\
$^2$National Research University Higher School of Economics, 101000 Moscow, Russia\\
}

\date{\today}
\begin{abstract}
We develop an explicitly gauge-invariant semiclassical approach to investigate nonlinear response of superconductors to monochromatic THz radiation. We demonstrate that in clean superconductors charge conservation forbids non-linear response to a uniform field. We apply our approach to quasi-two-dimensional films and obtain an explicit expression for the photoinduced current.  We find that the photoinduced current exhibits a strong dependence on polarization and incidence angle of the radiation. Our predictions may be directly verified in experiments with quasi-two-dimensional superconducting films.

\end{abstract}
\maketitle

\section{Introduction}
Recently, response of superconductors to electromagnetic field received a surge of attention following the advances in nonlinear terahertz (THz) spectroscopy \cite{kampfrath2013resonant}. Sufficiently intensive and coherent THz radiation allows to explore nonlinear response in a characteristic range of frequencies of order of the superconducting gap $\Delta_0$. Of particular interest are frequencies near $2\Delta_0$ where direct quasiparticle excitation becomes possible.

Superconductors are known to host a number of collective modes associated with variations of the complex superconducting order parameter $\Delta = |\Delta|{\rm e}^{i\varphi}$, see \cite{kulik}. The phase degree of freedom $\varphi$ is coupled to the electromagnetic field and is responsible for the Meissner effect, as well for the phenomenon of dissipationless current. Associated with phase is the superconducting plasma mode whose properties vary significantly depending on the temperature and the effective dimensionality of the sample. This mode has been extensively studied \cite{carlson1975,kulik,mooij} as it significantly affects linear response of superconductors. Another mode, usually referred to as the Higgs mode or Schmid mode \cite{schmid1968}, is associated with variations of the absolute value of the order parameter and has a characteristic frequency $\omega = 2\Delta_0$ . Unlike phase variations, in BCS superconductors this mode is decoupled from electromagnetic field in linear order which for long prevented its direct observation. 

In the last decade, the Higgs mode received extensive attention in the context of experiments \cite{matsunaga2013,matsunaga2014} where excitation of this mode serves as a possible explanation of the observed features of response at characteristic frequency $2\Delta_0$. A significant number of theoretical works which use different techniques has since been dedicated to the problem of nonlinear response of superconductors, especially to the third harmonic generation \cite{cea2016,murotani2019,tsuji2020,yang2021}. However, these works attribute the observed effects to different excitations: while some claim the Higgs mode to produce the largest contribution \cite{murotani2019,tsuji2020}, others find the contribution of density fluctuations dominant \cite{cea2016,yang2021}.

One of the challenges posed by the problem of nonlinear response is the gauge invariance. Within the standard diagrammatic approach, obtaining gauge-invariant response kernels requires accuracy even in the first order \cite{arseev2006}. In this paper, we aim to formulate a physically transparent approach where gauge invariance manifests itself outright and is identically satisfied in all orders of perturbation theory. We also apply our approach to calculate photoinduced current in quasi-two-dimensional superconducting films.

The structure of this paper is as follows. In the following section we define the system under our consideration. Then we specify the general formalism used in our analysis and formulate our approach. In the following sections we employ our approach to analyze nonlinear response of a quasi-two-dimensional film to microwave radiation. Finally, in the last section we discuss our findings and compare them to the results of other authors.

\section{System under consideration}

\begin{figure}
\begin{center}
\includegraphics[width=0.49\textwidth]{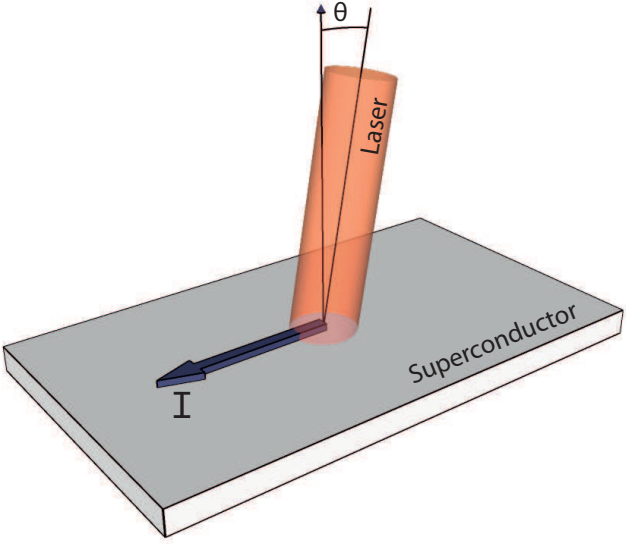}
\end{center}
\caption{Schematic depiction of the system under consideration. A large quasi-two-dimensional superconducting film subject to a monochromatic electromagnetic plane wave falling at an incidence angle $\theta$. 
}
\label{FIG:setup}
\end{figure}

Below we will consider a large superconducting film of constant thickness $d$ exposed to monochromatic radiation of frequency $\Omega$. We will be interested in frequencies of order $\Delta_0$ which usually lies in the THz range. We assume that the effects of size quantization may be neglected at the energy scale of $\Delta_0$ which usually holds for samples with $d$ of order of several decades of nanometres or larger. Effectively, this condition ensures that the electron dynamics inside the film may be considered three-dimensional. On the other hand, the dynamics of electric current and order parameter may be considered two-dimensional if $d$ is smaller than the London penetration depth $\lambda_L$ and the coherence length $\xi$ which usually are about from one to a few hundreds of nanometers. In this limit magnetic field can freely penetrate the film and, due to the large wavelength of radiation at these frequencies, all perturbations may be set constant along the transverse coordinate. However, dependence on coordinates in the plane of the film is allowed. The in-plane coordinate dependence of electromagnetic field is controlled by the incidence angle $\theta$ and is periodic with the wavevector $q = \Omega \cos\theta / c$. We will also assume our system to remain in thermodynamic equilibrium at a temperature $T$ below and not too close to the critical temperature $T_c$.

\section{General formalism}
The dynamics of an interacting BCS superconductor is described in terms of the complex superconducting order parameter field $\Delta = (\Delta_0 + \delta\Delta){\rm e}^{i\varphi}$ and electromagnetic field introduced via scalar and vector potentials $V$, $A$. In order to highlight gauge invariance, we perform a standard unitary transformation \cite{SCin1D,OGZB99} which leaves the order parameter real while incorporating its phase into electromagnetic potentials introducing gauge-invariant fields $\Phi = V + \dot{\varphi}/2e$, $\A = A - {c}\nabla\varphi/2e$.

For our problem, we chose to use Keldysh technique \cite{kamenevBook}. Therefore, for every field  $X$ we introduce components $X_{F}$, $X_{B}$ living on the forward and backward branches of the time contour, respectively. For convenience, we also perform Keldysh rotation to classical and quantum components $X_{cl,q} = (X_F \pm X_B)/2$. In order to somewhat compactify our notation, we also introduce matrices $\check{X} = \tau_0 X_{cl} + \tau_1 X_{q}$ where $\tau_0 = \begin{pmatrix}
1 & 0\\0 & 1
\end{pmatrix}$ and $\tau_1 = \begin{pmatrix}
0 &1\\1 & 0
\end{pmatrix}$ act in Keldysh space.

Within Keldysh technique, the dynamics of a BCS superconductor interacting with electromagnetic field is described by means of an effective action \cite{kamenevBook}
\begin{equation}
iS=iS_{EM}+iS_{\Delta}+\tr\, \ln\, \check{G}^{-1},\label{action}
\end{equation}
where
\begin{widetext}
\begin{align}
S_{EM} &= 4\int \frac{d\omega}{2\pi}\frac{d^3{k}}{(2\pi)^3}\Bigl\{
\Bigl(A_{cl}(\omega,k)-A_0(\omega,k)\Bigr)\frac{\omega^2/c^2-k^2}{8\pi}A_q(-\omega,-k)+ \Bigl(V_{cl}(\omega,k)-V_0(\omega,k)\Bigr)\,\frac{k^2}{8\pi} V_q\,(-\omega,-k)\Bigr\}
\end{align}
\end{widetext}
is the action of free electromagnetic field both inside and outside of the superconductor with $A_0, V_0$ describing the applied (laser) field. We assume the vector potential satisfies the Coulomb gauge condition $\nabla A = 0$. The next part
\begin{equation}
S_\Delta = -\frac{4}{g}\int dx dt\, \Delta_{cl}(x,t)\Delta_{q}(x,t)
\end{equation}
is the Hubbard-Stratonovich term for the order parameter with $g$ describing strength of the attractive BCS interaction. The last term $\tr\,\ln\,\check{G}^{-1}$ incorporates the contribution of electrons. The inverse electron Green function operator $\check{G}^{-1}$ has matrix structure in the tensor product of Nambu and Keldysh spaces (which we occasionally denote by indices $K,N$ over matrices if there is a need to clarify where they act), depends on collective fields and is given by 
\begin{align}
\check{G}^{-1} &= \check{1}^K \otimes \begin{pmatrix}
i\partial_t-\xi_{p-\frac{e}{c}\check{\A}}-e\check{\Phi} & \check{\Delta} \\
-\check{\Delta} & -i\partial_t-\xi_{p+\frac{e}{c}\check{\A}}-e\check{\Phi}
\end{pmatrix}^N\nn\\ 
& =\check{G}_0^{-1} - \check{X},\label{G_def}
\end{align}
where
\begin{align}
\check{G}_0 &= \begin{pmatrix}
G_0^R & G_0^K \\ 0 &G_0^A
\end{pmatrix}^K\\
{G}_0^{R,A}(\omega,p) & =\frac{1}{(\omega\pm i0)^2-(\Delta_0^2+\xi_p^2)}\begin{pmatrix}
\omega + \xi_p & \Delta_0 \\
-\Delta_0 & -\omega+\xi_p
\end{pmatrix}^N,\\
G_0^{K}(\omega, p) &= G_0^R(\omega ,p) F(\omega) - F(\omega) G_0^A (\omega,p)
\end{align}
is the bare (unperturbed) equilibrium Green function of a BCS superconductor with $\xi_p = (-i\nabla)^2/2m - \mu$, $\Delta_0$ - saddle point (BCS) value of the modulus of the order parameter and $F(\epsilon) = \tanh\frac{\epsilon}{2T}$ being the equilibrium electron distribution function, while
\begin{align}
\check{X} &= \left(e\check{\Phi} + \frac{e^2}{2mc^2}\check{\A}^2 \right)\sigma_0 - \delta\check{\Delta} i\sigma_2 - \frac{e}{2mc}\left\{\vec{p},\check{\vec{\A}}\right\}\sigma_3 \nn\\
&= \sigma_0 \check{X}_0 + i\sigma_2 \check{X}_2 + \sigma_3 \check{X}_3
\end{align}
includes perturbation terms. Matrices $\sigma_a$ are Pauli sigma matrices with $\sigma_0 = \hat{1}$ which act in Nambu space. Finally, full operator trace and multiplication operations involve both matrix multiplication and integration over internal space-time arguments. We will also use partial trace operations $\tr_{K,N}$ which only include tracing out matrix indices in Keldysh and Nambu spaces, respectively. One can see that the presented theory depends only on gauge-invariant potentials $\Phi$, $\mathcal{A}$, therefore, the gauge invariance is identically satisfied.

Average values of $X$ are given by a functional integral
\begin{equation}
\langle X \rangle = \int [DA\, DV\, D\Delta\, D\Delta^*] X_{cl}\, {\rm e}^{\ts iS}
\end{equation}
performed over both classical and quantum components of fields. The average electric current and charge density can be found as
\begin{align}
\bigl\langle \rho(x,t) \bigr\rangle &= \Bigl\langle \frac{i}{2}\frac{\delta}{\delta V_q(x,t)}\tr\,\ln\, \check{G}^{-1}\Bigr\rangle -  \rho_0^e = \nn\\\label{rho_def}
& = \Bigl\langle-\frac{ie}{2}\tr_N\left(G^K\right)(x,x,t,t)\Bigr\rangle  - \rho_0^e \\
\bigl\langle j(x,t) \bigr\rangle &= \Bigl\langle -\frac{ic}{2}\frac{\delta}{\delta A_q(x,t)}\tr\,\ln\, \check{G}^{-1}\Bigr\rangle = \nn\\\label{j_def}
& = \Bigl\langle-\frac{ie}{2m}\tr_N\left(\sigma_3\vec{p}\,G^K\right)(x,x,t,t)-\nn\\
&\ - \frac{e}{mc}\A_{cl}(x,t) (\rho(x,t)+ \rho^e_{0})\Bigr\rangle.
\end{align}
Here $\check{G}$ is the inverse operator to $\check{G}^{-1}$ given by (\ref{G_def}) and itself depends on fields while $\rho_0^e = \frac{ie}{2}\tr_N G_0^K(x,x,t,t)$ is the bare unperturbed electron charge density. Note that in (\ref{rho_def}) we need to subtract $\rho_0^e$ due to the presence of the static ionic background which exactly cancels the average electron density.

\section{Semiclassical approach}
Our problem as formulated above remains too complicated for analytical solution. Usually one expands action up to the second order in collective fields \cite{SCin1D,OGZB99} and treats subsequent nonlinear terms perturbatively. Alternatively, one can treat this problem semiclassically and expand the action only in quantum components of collective fields. Within this approach, the full nonlinearity of the classical dynamics is captured exactly while quantum fluctuations are considered small \cite{kamenevBook,radovskayaSemenov2021}. It is sufficient for our purposes to expand the action up to the first order in quantum components of all fields (this approximation will be justified below). Then integration over quantum components  yields
\begin{equation}
\bigl\langle X \bigr\rangle = \int [DA_{cl}\, DV_{cl}\, D\Delta_{cl}\, D\Delta^*_{cl}] X_{cl}\, \delta^f\left.\Bigl(
\frac{\delta S}{\delta X_{q}}
\Bigr)\right|_{X_q = 0}.
\end{equation}
All classical components assume values determined by equations of motion while quantum components remain zero. Taking variation of the action with respect to $V_q, A_q, \delta\Delta_q, \varphi_q$ consecutively, we arrive to the following set of equations
\begin{align}
&\frac{-\nabla^2}{4\pi}(V-V_0)=\rho, \label{eq_m1}\\
&\frac{-\nabla^2+\partial_t^2/c^2}{4\pi}(A-A_0)=\frac{1}{c}j, \label{eq_m2}\\
&\Delta=-\frac{ig}{4}\tr_N\Bigl( i\sigma_2 {G}^K\Bigr)(x,x,t,t), \label{eq_m3}\\
&0 = \dot{\rho}+\vec{\nabla}\vec{j}\label{eq_m4}
\end{align}
with $\rho, j$ given by Eqs.(\ref{rho_def})  and (\ref{j_def}). The first two are essentially Maxwell equations with electron density and current as sources. The third equation is the dynamic self-consistency equation. The fourths one is the continuity equation. It represents charge conservation and is needed to ensure gauge invariance of response functions. This set of equations together with 
\begin{equation}
\check{G}^{-1}[X] \check{G}[X] = \Bigl(
\check{G}_0^{-1} - \check{X}
\Bigr)\check{G}[X] = \check{1}\label{eq_m5}
\end{equation}
completely describes dynamics of a superconductor in our approximation. Let us notice that equation (\ref{eq_m5}) incorporates two independent equations. The retarded and advanced blocks describe changes of spectral properties of the electron system due to radiation while the Keldysh component accounts for the electron distribution function and may be used to derive a collisionless kinetic equation, similarly to ordinary plasma \cite{kamenevBook}.
Omission of higher orders of quantum components in action is a valid approximation under assumption that we can neglect quantum and thermal fluctuations. For typical laser intensities it is indeed justified. Nevertheless, the quantum nature of electron motion is incorporated into the theory in a similar manner it is within the commonly used self-consistent Born approximation which can be reproduced using this approach. One should also note that this approximation misses the collision integral. However, dissipation is still present due to the Landau damping. Another justification for the omission of the collision integral in the context of nonlinear response is the fact that superconductors exhibit little heating. Let us finally note that the resulting set of collisionless equations is analogous to Vlasov equations for plasma accompanied by the dynamic self-consistency equation and the continuity equation \cite{kamenevBook}.

In what follows we will consider electric field to be completely screened by the substrate and thus use a simplified version of Eq.\ref{eq_m1} in a form
\begin{equation}
C(V - V_0) = \rho \label{effective_coulomb_eq}
\end{equation}
where $C$ is the capacitance of the film per unit area \cite{SZ,zaikinBook}.

\section{Perturbative expansion}
Equations (\ref{eq_m2})-(\ref{effective_coulomb_eq}) are nonlinear. Therefore, in order to solve them, one generally needs to employ numeric methods. Luckily, the nonlinear response is usually sufficiently weak and can be treated perturbatively in external field $A_0$. We will seek the solution of Eqs.(\ref{eq_m2})-(\ref{effective_coulomb_eq}) taking $A_0$ as the expansion parameter. For the Green function we have
\begin{equation}
\check{G}=\check{G}_0 + \check{G}_0 \check{X}\check{G}_0 + \check{G}_0 \check{X}\check{G}_0 \check{X}\check{G}_0 + \dots
\end{equation}
As $X$ has matrix structure in Nambu space and since we need only $G^K(x,x,t,t)$ it is convenient to introduce kernels
\begin{align}
&{\chi}_{a,a_1,\dots,a_{n}}^{\,\,\,\Omega,\dots,\Omega_{n}}(p;q)=\frac{i}{2}\int\frac{d\omega}{2\pi}
\tr \Bigl(
\tau_1 \hat{\sigma}_{a} \check{G}_0(Q+Q_1+\ldots+Q_{n})\cdot\nn\\
\cdot &\hat{\sigma}_{a_1} \check{G}_0(Q+Q_2+\ldots+Q_{n})\cdot\ldots\cdot\hat{\sigma}_{a_n} \check{G}_0(Q)
\Bigr).\label{chi_def}
\end{align}
Here $Q$ includes both $\omega, p$ and $Q_i = (\Omega_i, q_i)$ stands for frequencies and momenta of perturbations. We assume $q_i$ to lie in two-dimensional plane of the film and $p$ to be 3-dimensional. The integral over $p$ is not performed at this stage as $X$ depends on the momentum through $\{p,\A\}$. With use of these kernels the perturbative expansion of right-hand sides of Eqs.(\ref{eq_m2})-(\ref{eq_m4}),(\ref{effective_coulomb_eq}) is given by

\begin{widetext}
\begin{align}
\rho(Q)
=&-e \sum\limits_p \sum\limits_{n=1}^\infty  \int\left[\prod\limits_{i=1}^{n}\frac{d^3 Q_i}{(2\pi)^3}\right] (2\pi)^3\delta(Q-(Q_1+\dots+Q_n))\cdot\chi_{0a_1\dots a_n}^{\Omega_1 \dots \Omega_n}(p;q) X_{a_1}(Q_1)\dots X_{a_n}(Q_n)
\\
\vec{j}(Q) =& - \frac{e}{m} \sum\limits_p \sum\limits_{n=1}^\infty  \int\left[\prod\limits_{i=1}^{n}\frac{d^3 Q_i}{(2\pi)^3}\right] (2\pi)^3 \delta(Q-(Q_1+\dots+Q_n))\cdot\left(\vec{p}+\frac{\vec{q}_1+\dots+\vec{q}_n}{2}\right)\chi_{3a_1\dots a_n}^{\Omega_1 \dots \Omega_n}(p;q) X_{a_1}(Q_1)\dots X_{a_n}(Q_n)\nn\\
& - \frac{e}{mc}\int\frac{d^3Q^\prime}{(2\pi)^3}\vec{\A}(Q-Q^\prime)\left[\rho(Q^\prime) + (2\pi)^3\delta(Q^\prime)\rho_0^e\right]
\\\label{current_expansion}
\Delta(Q)
=&-\frac{g}{2} \sum\limits_p \sum\limits_{n=0}^\infty  \int\left[\prod\limits_{i=1}^{n}\frac{d^3 Q_i}{(2\pi)^3}\right] (2\pi)^3\delta(Q-(Q_1+\dots+Q_n))\cdot\chi_{2a_1\dots a_n}^{\Omega_1 \dots \Omega_n}(p;q) X_{a_1}(Q_1)\dots X_{a_n}(Q_n).
\end{align}
\end{widetext}

Evaluation of kernels $\chi (p,q)$ with subsequent summation over momentum is a demanding task, especially for a dirty superconductor. Luckily, for our purposes it is not needed as we are interested in frequencies ranging from almost zero to several $\Delta$. At these frequencies typical light wavelengths are of order $10\mu m$ or (usually) larger. Hence, conditions $q \ll p_F$ and $q \ll 1/l_{el}$ where $p_F$ is the Fermi momentum and $l_{el}$ is the electron mean free path are well satisfied. Therefore, we can expand all kernels in momenta $q_i$ and consider only zeroth- and first-order (if needed) terms which are expressed in terms of kernels $\chi_{a\ldots a_n}(p)$ defined as $\chi_{a\ldots a_n}(p;q_1 = 0, \ldots, q_n = 0)$. This expansion is performed explicitly in Appendix.

Subsequent calculations can be further simplified if we take into account {Ward} identity
\begin{equation}
\hat{\sigma}_0\check{G}_0(\omega,p) - \check{G}_{0}(\omega+\Omega,p)\hat{\sigma}_0 = 
\check{G}_{0}(\omega+\Omega,p)\Bigl(
\Omega\hat{\sigma}_3
\Bigr)
\check{G}_0(\omega,p)\label{ward_1}.
\end{equation}
It allows us to express kernels $\chi(p)$ of order $n$ with index $3$ through kernels of order $n-1$ via
\begin{align}
&\chi\vphantom{|}_{a_0}\vphantom{|}_{a_1}^{\Omega_1}\vphantom{|}_{\dots}^{\dots}\vphantom{|}_{3}^{\Omega_i}\vphantom{|}_{\dots}^{\dots}\vphantom{|}_{a_n}^{\Omega_n} = \nn\\
&\left(\chi\vphantom{|}_{a_0}\vphantom{|}_{a_1}^{\Omega_1}\vphantom{|}_{\dots}^{\dots}\vphantom{|}_{,a_{i-1},}^{,\Omega_{i-1}+\Omega_{i},}\vphantom{|}_{a_{i+1}}^{\Omega_{i+1}}\vphantom{|}_{\dots}^{\dots}\vphantom{|}_{a_n}^{\Omega_n} - \chi\vphantom{|}_{a_0}\vphantom{|}_{a_1}^{\Omega_1}\vphantom{|}_{\dots}^{\dots}\vphantom{|}_{a_{i-1}}^{\Omega_{i-1}}\vphantom{|}_{,a_{i+1},}^{,\Omega_i+\Omega_{i+1},}\vphantom{|}_{\dots}^{\dots}\vphantom{|}_{a_n}^{\Omega_n}\right)/\Omega_i.\label{ward_2}
\end{align}
From this it immediately follows that kernels $\chi_{33\dots 3 a 3\dots 33}(p)$ where all indices, except maybe one, are equal to $3$, vanish identically. It is important to stress that such a claim is valid only in the limit of $q \rightarrow 0$ and $\Omega \neq 0$ since some of these kernels exhibit non-interchangeability of limits $q \rightarrow 0$ and $\Omega \rightarrow 0$. This fact ensures diamagnetic nature of current in the system which will be shown to have dramatic effect on the response.

It is also important to distinguish kernels that vanish in the presence of electron-hole (EH) symmetry. It its straightforward to verify (see Appendix) that kernels with even number of indices 0 are EH symmetric while those with odd number are EH asymmetric.

\section{Uniform case}
Now we are ready to investigate the nonlinear response itself starting with the simplest uniform case -- that is when the falling planewave is perpendicular to the film. Our observable of interest is electric current $j$. From rotational symmetry considerations the current includes only odd orders in $A_0$ while $\rho, \delta\Delta, \varphi$ have only even-order terms. As the incident wave is uniform along the film, so should also all the perturbations be. Now we take into account the continuity equation and immediately get $\rho(x,t) = 0$. 

The solution of Eqs.(\ref{eq_m1})-(\ref{eq_m5}) in now found trivially with the result
\begin{equation}
\vec{j} = -\frac{e}{mc}\vec{A}\rho_0^e.\label{uniform_case_current}
\end{equation}
For the third-order response we must carefully investigate all potentially contributing kernels in Eq.(\ref{current_expansion}). Such an analysis is carried out in Appendix. The result shows absence of third-order response in the uniform case as a result of the fact that the current in our system is purely diamagnetic due to the Ward identity (\ref{ward_2}).  A more thorough investigation shows that in the clean case this result should hold at all orders as a consequence of the identity 
\begin{equation}
\Bigl[\hat{\vec{j}}_{q = 0}, \hat{H}\Bigr] = 0 \label{current_commutator}
\end{equation}
which holds for uniform fields $\vec{A}(x,t) = \vec{A}(t)$. Therefore, within the adopted model with parabolic electron spectrum the nonlinear response occurs only for space-dependent perturbations (see, e.g,  \cite{cea2016},\cite{murotani2019} and references therein) or, equivalently, when the incident wave in not normal to the plane of the film.
In real materials with disorder the identity (\ref{current_commutator}) no longer holds, and optical response of the film may become non-trivial, see\cite{Schrieffer_Book}.In a real experiment, the third harmonic generation may also occur at the boundary of the sample due to various reasons, mainly due to the presence of defects. However, theoretical analysis of boundary effects lies beyond the scope of the present work.

\section{Non-uniform case. First-order response}
Now we will turn to the case of a non-uniform incident field which corresponds to the situation when the electromagnetic wave is not perpendicular to the film. Let us first investigate the first-order response within our approach. The first-order expressions for $\rho, j, \Delta$ are given by
\begin{align}
\rho^{(1)}(q) = & - e \sum\limits_p\left(\Phi^{q} \chi_{00}^{\Omega}(p;q) e - \chi_{02}^{\Omega}(p;q) \delta\Delta^{q}\right.\nn\\
&\left. - \frac{\chi_{03}^{\Omega}(p;q) e {\mathcal{A}}^{q}_{a_1}}{c m} \left({p}_{a_1} + \frac{{q}_{a_1}}{2}\right)\right)\nn\\
j^{(1)}_a (q) = &- \frac{\rho_0^e e {\mathcal{A}}^{q}_{a}}{c m} - \frac{e}{m} \sum\limits_p\left({p}_{a} + \frac{{q}_{a}}{2}\right) \left(\Phi^{q} \chi_{30}^{\Omega}(p;q) e \right.\nn\\
& \left.- \chi_{32}^{\Omega}(p;q) \delta\Delta^{q} - \frac{\chi_{33}^{\Omega}(p;q) e {\mathcal{A}}^{q}_{a_1}}{c m} \left({p}_{a_1} + \frac{{q}_{a_1}}{2}\right)\right)\nn\\
\Delta^{(1)}(q) = &- \frac{g}{2} \sum\limits_p\left(\Phi^{q} \chi_{20}^{\Omega}(p;q) e - \chi_{22}^{\Omega}(p;q) \delta\Delta^{q} \right.\nn\\
& \left.- \frac{\chi_{23}^{\Omega}(p;q) e {\mathcal{A}}^{q}_{a_1}}{c m} \left({p}_{a_1} + \frac{{q}_{a_1}}{2}\right)\right).\label{1_order_expansion}
\end{align}
A major simplification can be made if we take into account that $v_F \ll c$ which leads to $q \ll \Omega/v_F$ (in addition to the aforementioned condition $q \ll p_f, 1/l_{el}$. Being interested in frequencies in the microwave range, we see that all the kernels can be expanded in $q$ up to the lowest necessary order for which relatively simple expressions are available (see Appendix). Inclusion of terms of higher order in $q$ produces additional small parameter $\propto v_F/c$. We will also consider electromagnetic field generated by currents inside the film small compared to the incident laser field. It is justified as long as the film is thinner than the London penetration depth. This assumption allows us to set $A = A_0$ and completely discard the Maxwell equation (\ref{eq_m2}).

Taking into account these considerations, we now insert the expansion (\ref{1_order_expansion}) into equations \cref{eq_m3,eq_m4,effective_coulomb_eq} and find $V,\varphi$. Inserting these solutions back into Eq.(\ref{1_order_expansion}) yields the linear response current. In the simplest case of electron-hole symmetric material we get
\begin{equation}
\vec{j}^q = -\frac{\rho_0^e e}{mc}\left(
\vec{A}^q - \frac{\vec{q}(\vec{q}\vec{A}^q)}{q^2 - g(\Omega)\frac{\Omega^2}{v_0^2}}
\right)\label{j1_leading}
\end{equation}
where
\begin{equation}
v_0^2 = \frac{d \rho_0 e}{mC} \sim v_F^2 \frac{1}{\varepsilon}\left(\frac{d}{a_0}\right)^2
\end{equation}
is the plasma mode velocity \cite{mooij,zaikinBook} and
\begin{equation}
g(\Omega) = \frac{\sum\limits_p\chi_{00}^\Omega(p) e^2 d}{\sum\limits_p\chi_{00}^\Omega(p) e^2 d + C} \approx 1.
\end{equation}
Equation (\ref{j1_leading}) represents the first-order result for the electric current. The contribution of the Higgs mode is small due both electron-hole asymmetry and smallness of $q$ and is not included in (\ref{j1_leading}). However, even in electron-hole asymmetric materials the Higgs mode does not change much with its only effect being a renormalization of the plasma mode parameters. The expression for the current contains a perfect diamagnetic response term and a contribution from the plasma mode which has a pole at $\Omega \approx v_0 q$. If the plasma mode velocity satisfies $v_0 \ll c$, then we arrive to $\vec{j}^{(1)} \approx -\frac{\rho_0^e e}{mc} \left(\vec{A} - \frac{\vec{q}(\vec{q}\vec{A})}{q^2}\right)$. Note that the expression in brackets is not equal to $A^\perp $ since here $A$ and $q$ are projections of corresponding 3-dimensional wave amplitude and wave vector on the plane of the film.

\section{Non-uniform case. Photoinduced current}
Now let us proceed to the second order of the perturbative expansion. As $A(x) = A_q{\rm e}^{iqx} + A^*_q {\rm e}^{-iqx}$, in the second order we naturally get current response at wavevectors $\pm 2q, 0$. Here, we will consider specifically a stationary and uniform component of the current. In the regime of fluctuating superconductivity (at $T > T_c$) the effect was studied in \cite{boev}. Normally, in bulk crystals with inversion symmetry the second-order response is forbidden. In a thin film geometry, this is no longer the case as there exists a unique direction normal to the plane of the film. The current is allowed to flow only along the plane of the film and is then affected only by the in-plane components of $A$ which change with a spatial period defined by the in-plane wavevector $q$. The projection of $A$ is not orthogonal to $q$, and one may easily obtain a nonzero second-order current response. The appearance of the second-order response in such systems may be thought of as a boundary effect and may happen not only in low-dimensional geometry but also at the boundary of a bulk sample.

After a series of simplifications including decomposition in $q$ and angular integration, the resulting expression in the second order assumes form
\begin{align}
j_a = & - \frac{e}{c m}\left(\mathcal{A}^q_a \rho^{-q} + \mathcal{A}^{-q}_a\rho^q\right) + \text{ \it subleading terms}
\label{j(0)_expanded}
\end{align}
Subleading terms result from the $n = 2$ expansion of (\ref{current_expansion}) and are smaller by parameter $\propto ({v_F}/{c})^2 ({\Omega}/{\Delta})^2 $ (for details see Appendix),  unless the frequency of the incident wave is not too close to $2\Delta$ where resonant excitation of collective modes takes place. Omission of higher-order terms becomes no longer possible for $\Omega - 2\Delta \sim \Delta \left(\frac{v_F}{c}\right)^2$ where their contribution is of the same order. However, since all kernels become singular in this limit, the response itself becomes nonlinear. In this regime Eqs.(\ref{eq_m2})-(\ref{effective_coulomb_eq}) can no longer be treated perturbatively and require alternative solution methods.

Collecting the answer together using the already obtained first-order solution, for the stationary current density we get
\begin{equation}
\vec{j} = -\frac{eC}{m c^2 d}g(\Omega)\frac{(\vec{q}\vec{A}_q) \Omega}{q^2 - \frac{\Omega^2}{v_0^2} g(\Omega)}\left(
\vec{A}_{-q} - \frac{\vec{q}(\vec{q}\vec{A}_{-q}) }{q^2 - \frac{\Omega^2}{v_0^2} g(-\Omega)}
\right).\label{shift_current_result}
\end{equation}
In sufficiently thin films where $v_0 \ll c$ this answer reduces to a simple form $j = \frac{\rho_0 e^2}{(mc)^2 \Omega}{\Bigl(\vec{q}\vec{A}_q\Bigr)\vec{A}_{-q}}/$ valid not too close to $\Omega = 2\Delta$. It is interesting that this answer is almost identical to the standard result for 2D systems \cite{ivchenko2012} with the main difference being the dependence on light polarization.

\begin{figure}
\begin{center}
\includegraphics[width=0.49\textwidth]{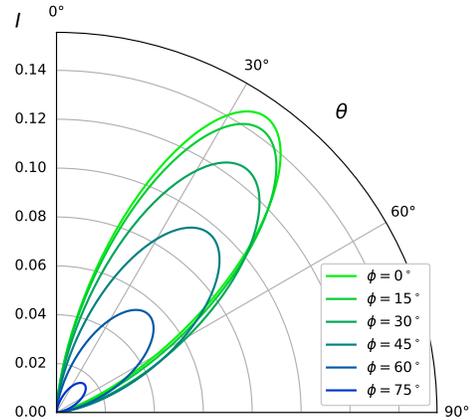}
\end{center}
\caption{Dependence of the second-order photoinduced current $I$ on the incidence angle $\theta$ for different angles $\phi$ in the limit $c \rightarrow \infty$ (see Eq. (\ref{current_angular_dependence})). At $\phi = 0$ the incident wave is transverse magnetically (TM) polarized while at $\phi = \pi/2$ the polarization is transverse electric (TE).}
\label{FIG:current}
\end{figure}

The current strongly depends both on the incidence angle of the light wave and on its polarization. The answer turns zero both in the limits of the wave being perpendicular and parallel to the plane of the superconducting film. It is also zero if the polarization of the wave is transverse electric (TE) for which $\vec{A} \perp \vec{q}$. Therefore, the largest values of the shift current are obtained for TM polarization at intermediate incident angles $\theta$. Let us now explore the full dependence of the photoinduced current on angle and polarization assuming that polarization components are given by $A^{TM} = A\cos\phi$ and $A^{TE} = A\sin\phi$. Taking into account that $\vec{q}$ is the projection of the full momentum of the incident wave, in the limit $v_0/c \rightarrow 0$ we get
\begin{equation}
|j|^2 \propto \cos^2\phi \sin^2\theta \cos^2\theta \Bigl(
\sin^2\phi + \cos^2 \phi \cos^2\theta
\Bigr), \label{current_angular_dependence}
\end{equation}
see FIG. \ref{FIG:current}. For a fixed linear polarization characterized by $\phi$ the maximal value of the current is obtained at incidence angle determined by $\cos^2\theta = \frac{1 + \cos^2\phi}{2 + \cos^2\phi}$.

It is important to stress that the contribution of the Higgs mode to the second-order response is suppressed by both approximate electron-hole symmetry and the ratio $v_F/c$ and may safely be omitted in our situation.

\section{Effective nonlinear action}
In the previous sections we identified leading first- and second-order terms in the perturbative expression for the current. Instead, it may be more convenient to carry out a perturbative expansion directly in the action (\ref{action}) where dominant terms may be determined using the same arguments as before. An effective action sufficient to reproduce the obtained results for the current is given by
\begin{align}
S_{eff} &= 2d \int\frac{d^3 k}{(2\pi)^3} \Bigl( - \tilde{\Phi}_{cl}(k) K_{\Phi \Phi}(k) \tilde{\Phi}_{q}(-k)  \nn\\
&- \frac{\rho_0^e e}{mc^2} \mathcal{A}_{cl}(k) \mathcal{A}_{q}(-k)
+ \delta\Delta_{cl}(k) K_{\Delta \Delta}(\Omega) \delta\Delta_{q}(-k) \nn\\
&+ \delta\Delta_{cl}(k) K_{\Delta \Phi} (\Omega) \tilde{\Phi}_{q}(-k)
+ \delta\Delta_{cl}(-k) K_{\Delta \Phi} (-\Omega) \tilde{\Phi}_{q}(k)
\Bigr) \nn\\
&+ 2\int dt\, d^2x\,  C V_{cl} V_q \nn\\
& - 2 \int dt\int d^3 x \frac{H_{cl}H_q}{4\pi}\label{nonlinear_action}
\end{align}
with $k$ encompassing both $\Omega$ and $q$, the new potential $\tilde{\Phi}$ given by $\tilde{\Phi}_{F,B} = \Phi_{F,B} + \frac{e}{2 mc^2} \mathcal{A}_{F,B}^2 $, $\tilde{\Phi}_{cl,q} = (\tilde{\Phi}_{F} \pm \tilde{\Phi}_B)/2$, and the kernels being
\begin{align*}
K_{\Phi \Phi}(\Omega) &= e\sum\limits_p \chi_{00}^{\Omega}(p)\\
K_{\Delta \Phi}(\Omega) &= \sum\limits_p \chi_{20}^{\Omega}(p)\\
K_{\Delta \Delta}(\Omega) &= -\frac{4}{g} + \sum\limits_p \chi_{22}^{\Omega}(p).
\end{align*}
Nonlinearity in Eq.(\ref{nonlinear_action}) is introduced through $\tilde{\Phi}$ which contains $\mathcal{A}^2$. It may be shown that all other nonlinear terms (up to the 3-d order) only generate smaller corrections to the final expression for the current as they contain higher orders of $q$. Within our approximations, this action provides a complete classical description of our system generating all equations of motion. It can be used to evaluate other nonlinear effects, such as generation of the second and third harmonics. However, one should take into account that the action (\ref{nonlinear_action}) contains terms relevant specifically for current response while different types of response may be more sensitive to other terms.

Comparing the obtained action (\ref{nonlinear_action}) with \cite{cea2016}, one finds that the authors derive a quite similar effective action for uniform perturbations. Their action taken in the limit of parabolic electron dispersion coincides with ours if we identify ``density fluctuations'' $\rho$ in \cite{cea2016} with $V$, take $q \rightarrow 0$ and consider screening in the substrate in this limit negligible (i.e., set $C = 0$). The differences originate both from considering different geometries and from the fact that in our case of parabolic dispersion uniform response is forbidden, so the phase gradient must be accounted for. As our analysis shows, the variable $\varphi$ is always crucial for obtaining gauge-invariant response functions.

It is also necessary to mention that the effective action (\ref{nonlinear_action}) is a simple nonlinear extension of the gaussian effective action obtained in \cite{SCin1D,OGZB99}. In these works, all the linear response kernels were evaluated in the dirty limit for arbitrary frequencies which serves a good starting point for further investigation of nonlinear microwave response in the dirty limit.

Finally, the effective action (\ref{nonlinear_action}) can be easily modified in order to incorporate quantum and fluctuation effects. For this purpose it is sufficient to introduce terms $\propto X_q X_q$ while the coefficients may be restored using the fluctuation-dissipation theorem.


\section{Results and Discussion}
In this paper we investigated nonlinear response of a thin superconducting film in electromagnetic field. We developed an approach based on equations of motion for the collective fields in the spirit of self-consistent Vlasov equations used to describe the dynamics of interacting plasma. The resulting set of equations (\ref{effective_coulomb_eq}), (\ref{eq_m2})-(\ref{eq_m4}) along with (\ref{eq_m5}) governs the dynamics of electromagnetic field and order parameter and has a clear physical interpretation and includes Coulomb equations, self-consistency equation, continuity equation as well as the equation for the Green function.  In contrast with the more widely used diagrammatic approach where in order to achieve gauge invariance one needs to carefully sum various diagrammatic contributions \cite{arseev2006}, within our approach the charge conservation manifests itself right away: the equation of motion for the superconducting phase coincides with the continuity equation for electric charge.

Charge conservation is particularly important in the context of electromagnetic response of a superconductor due to the diamagnetic nature of the superconducting current. At least for pure superconductors, as long as the external field varies slowly in space ($q \ll \Omega/v_F$), the paramagnetic response is very weak and vanishes identically for uniform fields in several lowest orders (see Eq. (\ref{ward_2}) and the paragraph below). Therefore, the current is given by a simple expression $\vec{j} = -\frac{e}{mc}\rho (\vec{A}-\frac{c}{2e} \vec{\nabla}\varphi)$. In a uniform field, $\varphi(x)$ is uniform in space while the density assumes its equilibrium value due to the charge conservation. Hence, only the linear term $\vec{j} = -\frac{e}{mc}\rho_0 \vec{A}$ survives reproducing a well-known classical answer \cite{schriefferBook}. 

It is instructive to compare this conclusion to the result of \cite{cea2016} where authors studied third harmonic generation in bulk superconductors. They find that the third-order current is determined predominantly by the diamagnetic term with a (usually) small correction produced by the amplitude mode, in agreement with our results.
A closer examination of the response kernels obtained in \cite{cea2016} shows that after renormalization by phase degrees of freedom they obtain a correction which cancels the kernels out exactly in the case of parabolic electron dispersion leaving the third harmonic zero. This again highlights the role of the phase mode.

The crucial role of scalar fields $\varphi$ and $V$ in the context of nonlinear response was also emphasized in \cite{yang2021} where authors showed how their dynamics cancels out density response from other channels. The authors also study the problem of third harmonic generation. According to their results, the main contribution to the third-order response (at least, in absence of voltage applied to the film) comes from the term $S_{AA\Delta} \propto K^{\Omega\Omega}_{AA\Delta}\mathcal{A}^2(\Omega)\delta\Delta(-2\Omega)$ which generates current $j(3\Omega) \propto A(\Omega) \frac{\Bigl(K^{\Omega\Omega}_{AA\Delta}(2\Omega)\Bigr)^2}{K_{\Delta\Delta}} A^2(\Omega)$. They claim it to persist in the limit of $q \rightarrow 0$ and provide an expression for the necessary kernel in the limit $\lim\limits_{\Omega \rightarrow 0} \Bigl(\lim\limits_{q \rightarrow 0} K_{AA\Delta} \Bigr)$ which remains finite. We find the latter statement erroneous since the identities (\ref{ward_1}) and (\ref{ward_2}) leave this kernel zero at $q = 0$. One may obtain a nonzero expression for this kernel in the opposite limit $\lim\limits_{q\rightarrow 0}\lim\limits_{\Omega\rightarrow 0}$ which is relevant in the context of response to non-uniform static fields, but not in our situation.
Such behavior is an example of a well-known non-analyticity which many response kernels exhibit at $q,\Omega \rightarrow 0$ with the most common example being the Linhard function for the polarization operator \cite{lindhard1954,altlandBook}. In the context of third harmonic generation in superconductors, we find again that in our situation the evidence for the dominance of the Higgs mode is insufficient.

Our results demonstrate that in clean isotropic superconducting films second- and third-order responses to a uniform microwave field vanish. This implies  strong dependence of the response on the incidence angle of the microwave, as well as on its polarization. Absence of response at $q=0$ introduces necessity to expand the response kernels in momenta which in its turn makes the resulting current small by additional parameter $v_F/c$. This additional smallness provides an edge to thin films as they allow angle-dependent nonlinear response already in the second order. 
In our analysis, we focus our attention on the shift current -- a stationary or, in practice, slowly varying, component of the photoinduced current, for which we derive an explicit analytical expression given in Eq. (\ref{shift_current_result}). It is proportional to the total carrier density and shows weak dependence on other parameters of the superconducting material or the substrate, as long as the superconducting film is properly insulated and the velocity of the plasma mode remains much smaller than $c$. The frequency dependence is simple with $|j| \propto I/\Omega^2 $ outside of a narrow range of frequencies near $\Omega = 2\Delta $ where the current displays non-analyticity in all orders while a perturbative calculation becomes impossible. In this case Eqs.(\ref{eq_m1})-(\ref{eq_m5}) must be solved using alternative methods, for example, numerical ones. It should be stressed that the resonance at $\Omega = 2\Delta$ occurs in our analysis due to the excitation of the plasma mode while the contribution of the Higgs mode is negligibly small.

It is important to discuss the range of validity of our results. They are based on two premises: the diamagnetic nature of current in superconductors and the conservation of electric charge. The first one is a consequence of the identity (\ref{ward_1}) which holds for pure superconductors and leaves zero all the paramagnetic response kernels in the limit $q \rightarrow 0$, $\Omega \neq 0$ at any temperature. In disordered systems such a simple relation no longer exists and paramagnetic response may arise. 
Moreover, several studies claim paramagnetic response of higher orders to be much stronger in the presence of disorder \cite{murotani2019,tsuji2020}. Thus, the disordered case requires separate treatment and will be worked out elsewhere.

The charge conservation is another issue which should be addressed here as it requires the film to be completely insulated from other electron reservoirs. It might be especially difficult to achieve in DC measurements which usually require electrodes connected to the film and involve larger timescales. If charge is allowed to be transferred between the film and the environment, then the charge conservation for the film alone holds no more, and response at $q = 0$ becomes possible, although in this case it becomes largely dependent on the environment and cannot be considered a property of superconductor alone, but rather that of the whole setup.

Finally, even if the film is sufficiently pure and is properly insulated, it may still demonstrate angle-independent third-order response due to the boundary effects. As the bulk third-order response is relatively weak due to its dependence on the angle, the boundary might hide it in a given sample. Therefore, a film of sufficiently large area may be required. Within our approach, it is possible to account for boundary effects due to the fact that the effective action (\ref{nonlinear_action}) only includes spatial derivatives of first order. In order to do this, boundary conditions on charge and current must be imposed.

Our findings may be tested in experiments with quasi-two-dimensional films.

We would like to thank P.I. Arseev for helpful discussions. We acknowledge support by RFBR grant № 18-29-20033.

\bibliography{NL_response}

\section{Appendix}
\subsection{Kernels}
Evaluating integrals in Eq.(\ref{chi_def}) at $q = 0$, one obtains
\begin{align}
\sum\limits_p\chi_{00}^\Omega(p) &= \frac{2\nu_0}{y\sqrt{1 - y^2}}\arctan\frac{y}{\sqrt{1 - y^2}}\\
\sum\limits_p\chi_{003}^{0\Omega}(p) &= \frac{\nu_0}{2\Delta y}\left(
-1 + \frac{\sqrt{1-y^2}}{y}\arctan \frac{y}{\sqrt{1 - y^2}}
\right)\\
\sum\limits_p\chi_{22}^\Omega(p) &= (y^2 - 1)\sum\limits_p\chi_{00}^\Omega(p)
\end{align}
where $T \rightarrow 0 $ and $y = (\Omega + i0)/2\Delta$.
\subsection{Spectral decomposition}
Any retarded/advanced propagator $f^{R,A}$ can be expressed via Lehman representation
\begin{equation}
f^{R,A}(\omega) = \int\limits_{-\infty}^\infty dz \frac{J_f(z)}{\omega\pm i0 -z},
\end{equation}
where 
\begin{equation}
J_f(\omega) = \frac{i}{2 \pi} \Bigl(f^R(\omega)-f^A(\omega)\Bigr).
\end{equation}
For bare equilibrium Green function of a BCS superconductor
\begin{equation}
\hat{J}_p(\omega) = 
J_p(\omega)\cdot \hat{M}_p^{(N)}(\omega),
\end{equation}
with
\begin{align}
J_p(\omega) & = \frac{1}{2\varepsilon_p}\Bigl(
\delta(\omega-\varepsilon_p) - \delta(\omega+\varepsilon_p)
\Bigr)\\
\hat{M}_p(\omega) & = (\xi_p \hat{\sigma}_0 + \Delta i\hat{\sigma}_2 + \omega\hat{\sigma}_3)^N.
\end{align}
Accordingly,
\begin{align}
\check{G}_0(\omega,p) &= \int\limits_{-\infty}^\infty dz \begin{pmatrix}
\ds \frac{1}{\omega + i0 -z} & -2\pi i \delta(\omega - z)F_\omega \\
0 & \ds \frac{1}{\omega - i0 -z}
\end{pmatrix}^K J_p(z) \hat{M}_p(z) \\
&= \begin{pmatrix}
S^R(\omega) & \Bigl[S^R(\omega) - S^A(\omega)\Bigr]F_\omega \\
0 & S^A(\omega)
\end{pmatrix}^{K} \otimes \hat{M}_p(\omega), \\
S^{R,A}(\omega) &= \frac{1}{(\omega\pm i0)^2 -\varepsilon_p^2}.
\end{align}

Spectral decomposition can be used in regularization of kernels or to analyze their non-trivial limits. Separation of the Nambu space structure in matrix $\hat{M}$ is useful for determining some properties of kernels.

\subsection{EH symmetry of kernels $\chi(p)$}
Let us consider how kernel $\chi_{aa_1\dots a_n}(p)$ changes upon substitution $\xi_p \rightarrow -\xi_p$. We notice that kernel's symmetry is determined by the symmetry of trace 
\begin{equation}
Tr = \tr_N \Bigl(\sigma_{a}\hat{M}(\omega+...)\sigma_{a_1} \hat{M}(...)\dots \sigma_{a_n}\hat{M}(\omega)
\Bigr).
\end{equation}
Another observation is $
\hat{M}(\xi \rightarrow -\xi) = -\sigma_1\hat{M}\sigma_1
$. Inserting this relation into the just mentioned trace $Tr$ and performing a cyclic permutation, we get the same expression multiplied by $(-1)^{n_0}$ where $n_0$ is the number of indices $0$ in the kernel.

\subsection{Momentum expansion of kernels $\chi(p;q)$}
Let us use identity
\begin{equation}
\frac{\partial G_0(\omega,p)}{\partial_p} = - G_0(\omega,p) \frac{\partial G_0^{-1}(\omega,p)}{\partial p} G_0(\omega,p)
\end{equation}
Therefore,
\begin{equation}
G_0(\omega,p+q) \approx G_0(\omega,p) - \vec{v}\vec{q}\,G_0(\omega,p) G_0(\omega,p).
\end{equation}
Inserting the last equality in the definition of kernels $\chi(p;q)$, we have
\begin{dmath}
\chi_{a, a_1,\dots, a_n}^{\Omega_1,\dots, \Omega_n}(p;q_1,\ldots, q_n) - \chi_{a, a_1,\dots, a_n}^{\Omega_1,\dots, \Omega_n}(p) \approx - \vec{v}(\vec{q_1} + \ldots + \vec{q}_n) \chi_{a, 0, a_1, \ldots, a_n}^{\ 0, \Omega_1, \dots, \Omega_n}(p)  -
 -\vec{v}(\vec{q_2} + \ldots + \vec{q}_n) \chi_{a,  a_1, 0 , a_2 \ldots, a_n}^{\  \Omega_1, 0 ,\Omega_2, \dots, \Omega_n}(p) - \ldots
- \vec{v} \vec{q}_n \chi_{a, a_1, \ldots, a_{n-1} ,0, a_n}^{\  \Omega_1, \dots, \Omega_{n-1}, 0 ,\Omega_n}(p).
\end{dmath}
In case of extreme necessity such decomposition may be carried out in higher orders.

\subsection{Third-order response in the uniform case}
In the uniform case, due to the charge conservation the diamagnetic contribution to the current is given by (\ref{uniform_case_current}). Hence, we only need to consider paramagnetic contributions. Symmetry considerations show that from all the collective fields only $\mathcal{A}$ obtains first-order corrections. From the third-order expansion of the current (\ref{current_expansion}) only a single term with $\chi_{3333}$ remains relevant. However, the Ward identity (\ref{ward_2}) leaves it identically zero. From the second-order expansion angular integration leaves terms containing kernels with two indices $3$, such as $\chi_{303}$ which again vanish by virtue of the Ward identity (\ref{ward_2}). In the first order all paramagnetic terms are trivially zero. 

As a result, the current does not acquire second- and third-order corrections in the uniform case as a result of the charge conservation and the Ward identity (\ref{ward_1}) which cancels all paramagnetic contributions. Therefore, Eq.(\ref{uniform_case_current}) is correct at least up to the fifth order.

\subsection{Subleading terms of photoinduced current}

The subleading terms of the photoinduced current originate from the $n = 2$ order of Eq.(\ref{current_expansion}). They vanish at $q = 0$, therefore, require decomposition in $q$. As the result, for them we have
\begin{align}
&j_{a}^{subl} = - \frac{e}{m} \sum\limits_p p^2 \times\nn\\
&\left( \frac{\Phi^{-q} \chi_{3002}^{-\Omega0\Omega} \delta\Delta^{q} e}{3 m}  {q}_{a} - \frac{\Phi^{-q} \chi_{3200}^{\Omega0-\Omega} \delta\Delta^{q} e}{3 m}  {q}_{a}\right. \nn\\
& - \frac{\Phi^{q} \chi_{3002}^{\Omega0-\Omega} \delta\Delta^{-q} e}{3 m}  {q}_{a} + \frac{\Phi^{q} \chi_{3200}^{-\Omega0\Omega} \delta\Delta^{-q} e}{3 m}  {q}_{a} \nn\\
& + \frac{\Phi^{-q} \chi_{3003}^{-\Omega0\Omega} e^{2} {\mathcal{A}}^{q}_{a_2} {q}_{a_2} {q}_{a}}{6 c m^{2}}  + \frac{\Phi^{-q} \chi_{3300}^{\Omega0-\Omega} e^{2} {\mathcal{A}}^{q}_{a_1} {q}_{a_1} {q}_{a}}{6 c m^{2}} \nn\\
&  + \frac{\Phi^{q} \chi_{3003}^{\Omega0-\Omega} e^{2} {\mathcal{A}}^{-q}_{a_2} {q}_{a_2} {q}_{a}}{6 c m^{2}}  + \frac{\Phi^{q} \chi_{3300}^{-\Omega0\Omega} e^{2} {\mathcal{A}}^{-q}_{a_1} {q}_{a_1} {q}_{a}}{6 c m^{2}} \nn\\
& - \frac{\chi_{3203}^{-\Omega0\Omega} \delta\Delta^{-q} e {\mathcal{A}}^{q}_{a_2} {q}_{a_2} {q}_{a}}{6 c m^{2}}  - \frac{\chi_{3203}^{\Omega0-\Omega} \delta\Delta^{q} e {\mathcal{A}}^{-q}_{a_2} {q}_{a_2} {q}_{a}}{6 c m^{2}}\nn\\
&\left. - \frac{\chi_{3302}^{-\Omega0\Omega} \delta\Delta^{q} e {\mathcal{A}}^{-q}_{a_1} {q}_{a_1} {q}_{a}}{6 c m^{2}}  - \frac{\chi_{3302}^{\Omega0-\Omega} \delta\Delta^{-q} e {\mathcal{A}}^{q}_{a_1} {q}_{a_1} {q}_{a}}{6 c m^{2}} \right)\label{subl_j_0_terms}
\end{align}
They all contain kernels of high orders and are proportional to $q^3$. As compared to the diamagnetic contribution $\propto \rho \mathcal{A}$, they are small. Therefore, the photoinduced current is predominantly determined by diamagnetic terms associated with charge density variations.

Let us investigate the range of validity of the above statement. For this purpose we need to compare similar terms from Eq.(\ref{j(0)_expanded}) with those from Eq.(\ref{subl_j_0_terms}). For this purpose we need explicit expressions for the relevant kernels. For $(\Phi\A)$-terms at $T \rightarrow 0 $ the relevant kernels read
\begin{align}
{\chi}_{00}^\Omega(p) &= - \frac{4  \Delta^{2}}{\varepsilon_p \left(\Omega_1 - 2 \varepsilon_p\right) \left(\Omega + 2 \varepsilon_p\right)}\\
\chi_{3300}^{\Omega 0 -\Omega}(p) &= \chi_{3003}^{-\Omega 0 \Omega} = \frac{ \Delta^{2} \left(\Omega^{2} + 4 \varepsilon_{p}^{2}\right)}{\varepsilon_{p}^{3} \left(\Omega - 2 \varepsilon_{p}\right)^{2} \left(\Omega + 2 \varepsilon_{p}\right)^{2}}
\end{align}
At $\Omega \rightarrow 0$ we have 
\begin{align}
&\sum\limits_p\chi_{00}(p) = 2\nu_0,\\ 
&\sum\limits_p p^2\Bigl(\chi_{3300}^{\Omega 0 -\Omega}(p) + \chi_{3003}^{-\Omega 0 \Omega}(p)\Bigr) \approx \frac{8}{3}\frac{\nu_0 p_F^2}{\Delta^2}.
\end{align}
Then the leading contribution $j^{lead}_{\Phi\A}$ and the subleading term $j^{subl}_{\Phi\A}$ proportional to $\Phi^{-q}\A^{q}$ are given by
\begin{align}
\vec{j}^{lead}_{\Phi\A} & = \frac{e^3}{mc} 2\nu_0 \Phi^{-q}\vec{\A}^{q},\\
\vec{j}^{subl}_{\Phi\A} & = -\frac{e^3}{6m^3 c} \frac{8}{3}\frac{\nu_0 p_F^2}{\Delta^2}\Phi^{-q} \bigl( \vec{\A}^{q}\vec{q} \bigr)\vec{q}.
\end{align}
It is now straightforward to verify that the subleading contribution is indeed small by parameter $\propto ({v_F}/{c})^2 ({\Omega}/{\Delta})^2 $. 

The situation changes for frequencies $\Omega$ close to $2\Delta$ where kernels cease to be analytic functions of $\Omega$. For $(\Omega - 2\Delta)/\Delta = \varepsilon \ll1$ in the leading order we have
\begin{align}
&\sum\limits_p\chi_{00}(p) \approx \frac{\nu_0 \pi}{\sqrt{\varepsilon}},\\ 
&\sum\limits_p p^2\Bigl(\chi_{3300}^{\Omega 0 -\Omega}(p) + \chi_{3003}^{-\Omega 0 \Omega}(p)\Bigr) \approx \frac{\nu_0 \pi p_F^2}{4\Delta^2 \varepsilon^{3/2}}.
\end{align}
Now we have $j^{subl}_{\Phi\A}/j^{lead}_{\Phi\A} \sim \ds\frac{1}{12\varepsilon} \left(\frac{v_F}{c}\right)^2$. Hence, omission of higher-order kernels is unjustified in a narrow range of frequencies $\Omega - 2\Delta \sim \Delta \left(\frac{v_F}{c}\right)^2$. We also need to point out that in this range the response should become highly nonlinear due to the more singular behavior of higher-order kernels.


\end{document}